\documentclass[aps,prl,twocolumn,superscriptaddress,preprintnumbers]{revtex4}

\usepackage{epsf,epsfig,amsmath,amssymb,amsfonts}
\usepackage{color}
\usepackage[thinlines]{easytable}
\usepackage{slashed, comment}
\usepackage{hyperref}

\PassOptionsToPackage{table}{xcolor}
\hypersetup{ linktoc=all,
    colorlinks, linkcolor={blue}, 
    citecolor={red}, urlcolor={darkpink}
}

\graphicspath{{Images/}}

\definecolor{light gray}{RGB}{220,220,220}
\definecolor{dark purple}{RGB}{108,0,217}
\definecolor{pink}{RGB}{190,20,100}
\definecolor{orang}{RGB}{193,63,0}
\definecolor{green}{RGB}{11,98,17}
\definecolor{darkpink}{RGB}{153,0,76}
\definecolor{bluegreen}{RGB}{0,102,102}
\definecolor{greenlagan}{RGB}{0,102,0}
\definecolor{redgreen}{RGB}{102,102,0}
\definecolor{Redgreen}{RGB}{153,76,0}
\definecolor{vividviolet}{rgb}{0.62, 0.0, 1.0}
\definecolor{amaranth}{rgb}{0.9, 0.17, 0.31}
\definecolor{palatinateblue}{rgb}{0.15, 0.23, 0.89}
\definecolor{brightpink}{rgb}{1.0, 0.0, 0.5}
\definecolor{cornflowerblue}{rgb}{0.39, 0.58, 0.93}
\definecolor{deepcarminepink}{rgb}{0.94, 0.19, 0.22}
\definecolor{radicalred}{rgb}{1.0, 0.21, 0.37}

\newsavebox{\ns}
\newsavebox{\dbrane}

\def\be{\begin{equation}}
\def\ee{\end{equation}}
\def\bea{\begin{eqnarray}}
\def\eea{\end{eqnarray}}

\def\Dslash{\,\,{\raise.15ex\hbox{/}\mkern-12mu D}}
\def\Dbarslash{\,\,{\raise.15ex\hbox{/}\mkern-12mu {\bar D}}}
\def\delslash{\,\,{\raise.15ex\hbox{/}\mkern-9mu \partial}}
\def\delbarslash{\,\,{\raise.15ex\hbox{/}\mkern-9mu {\bar\partial}}}
\def\pslash{\,\,{\raise.15ex\hbox{/}\mkern-9mu p}}
\def\calDslash{\,\,{\raise.15ex\hbox{/}\mkern-12mu {\cal D}}}

\newcommand\diff{\mbox{d}}

\newcommand{\nn}{\nonumber \\}

\newcommand{\dd}{\diff}

\allowdisplaybreaks

\begin{document}

\title{Is there an early Universe solution to Hubble tension?}

\vskip 1 cm

\author{C. Krishnan}\email{chethan.krishnan@gmail.com}
\affiliation{Center for High Energy Physics, Indian Institute of Science, Bangalore 560012, India}
 \author{E. \'O Colg\'ain}\email{ocolgain.eoin@apctp.org}
 \affiliation{Asia Pacific Center for Theoretical Physics, Postech, Pohang 37673, Korea}
\affiliation{Department of Physics, Postech, Pohang 37673, Korea}
\author{Ruchika}\email{ruchika@ctp-jamia.res.in}
\affiliation{Centre for Theoretical Physics, Jamia Millia Islamia, New Delhi-110025, India}
\author{A. A. Sen}\email{aasen@jmi.ac.in}
\affiliation{Centre for Theoretical Physics, Jamia Millia Islamia, New Delhi-110025, India}
\author{M. M. Sheikh-Jabbari}\email{jabbari@theory.ipm.ac.ir}
\affiliation{School of Physics, Institute for Research in Fundamental Sciences (IPM), P.O.Box 19395-5531, Tehran, Iran}
\affiliation{The Abdus Salam ICTP, Strada Costiera 11, 34151, Trieste, Italy}
\author{T. Yang}\email{tao.yang@apctp.org}
\affiliation{Asia Pacific Center for Theoretical Physics, Postech, Pohang 37673, Korea}

\begin{abstract}
We consider a low redshift $(z<0.7)$ cosmological dataset comprising megamasers, cosmic chronometers, type Ia SNe and BAO, which we bin according to their redshift. For each bin, we read the value of $H_0$ by fitting directly to the flat $\Lambda$CDM model. Doing so, we find that $H_0$ descends with redshift, allowing one to fit a line with a \textit{non-zero} slope of statistical significance $2.1 \, \sigma$. Our analysis rests on the use of cosmic chronometers to break a degeneracy in BAO data and it will be imperative to revisit this feature as data improves. Nevertheless, our results provide the first independent indication of the descending trend reported by the H0LiCOW collaboration. If substantiated going forward, early Universe solutions to the Hubble tension will struggle explaining this trend. 
\noindent 

\end{abstract}

\maketitle

\setcounter{equation}{0}

\section{Introduction} \label{Introduction}

Driven by successive results \cite{Riess:2019cxk, Wong:2019kwg, Pesce:2020xfe} favouring a higher value of the Hubble constant $H_0$, cosmology is in a state of flux. Remarkably, independent local determinations of $H_0$ based on Cepheids/Type Ia Supernovae (SNe) (SH0ES) \cite{Riess:2019cxk}, strongly-lensed quasar time delay (H0LiCOW) \cite{Wong:2019kwg} and water megamasers (The Megamaser Cosmology Project) \cite{Pesce:2020xfe} all appear to be converging to an {overall} result that is discrepant at the $\sim 5 \, \sigma$ level with the lower value reported by the Planck mission based on the cosmological model $\Lambda$CDM \cite{Aghanim:2018eyx} (see \cite{Verde:2019ivm} for a review). Moreover, simple physically motivated changes in dark energy seem not to significantly alter the local distance ladder  calibration by Cepheids \cite{Dhawan:2020xmp}.

{Given the tension, it is imperative to address systematics, and calibration is a good place to start.} Replacing Cepheids with Tip of the Red Giant Branch as a calibrator for Type Ia SNe, the Carnegie-Chicago Hubble Program have found an intermediate value \cite{Freedman:2019jwv}, which is about $1.2 \, \sigma$  away from the Planck value and $1.7 \, \sigma$, $1.3 \, \sigma$, $1.2 \, \sigma$ from SH0ES, H0LiCOW and Megamaser Cosmology Project, respectively. The result is the subject of an ongoing dispute \cite{Yuan:2019npk, Freedman:2020dne}. Nevertheless, it has also recently been demonstrated that a combination of low-redshift cosmological data also favour a central value $H_0 \sim 70$ km s$^{-1}$ Mpc$^{-1}$ that is within $2 \, \sigma$ of all experiments \cite{Dutta:2019pio}.

{Evidently some soul searching is required to convince ourselves that Hubble tension indeed implies deviations from $\Lambda$CDM. Among various  ideas put forward to address the tension, the early dark energy (EDE) proposal \cite{Poulin:2018cxd} has attracted a lot of attention. The idea is to retain $\Lambda$CDM for the late-time cosmology, but reduce the sound horizon radius at drag epoch $r_d$ by turning on a cosmological constant at early times ($z\gtrsim 3000$), which  dilutes away like radiation or faster at later times. A similar reduction in $r_d$ may be achieved through other mechanisms \cite{Kreisch:2019yzn, Agrawal:2019lmo,Bernal:2016gxb, Mortsell:2018mfj, Escudero:2019gvw, Alcaniz:2019kah}.

Here, motivated by a recent H0LiCOW observation that $H_0$ decreases with lens redshift \cite{Wong:2019kwg}, we take a closer look at, or alternatively, \textit{fine-grain} the dataset favouring an intermediate value \cite{Dutta:2019pio} to see if it hides a similar feature.  While the H0LiCOW result is of low statistical significance at $1.9 \, \sigma$ (reduced to $1.7 \, \sigma$ with the inclusion of DES J0408-5354 \cite{Shajib:2019toy}), importantly there is no indication that the trend is due to unaccounted systematics \cite{Millon:2019slk}. Note, changing the cosmology from $\Lambda$CDM to $w$CDM inflates the errors \cite{Wong:2019kwg}, so if real, this is a model-dependent trend, providing {a potentially new diagnostic for Hubble tension}. 

Concretely, our fine-graining process involves binning the Dutta et al. \cite{Dutta:2019pio}  dataset by redshift. This analysis has difficulties stemming from a decreasing quality of data as one increases redshift. Therefore, we restrict our attention to redshift $ z \leq 0.7$, more or less the range of interest for H0LiCOW, and adopt non-uniform bin sizes to identify the appropriate redshift favoured by data.

As expected, we recover the $H_0$ value of \cite{Dutta:2019pio} from data in the range $0.002 < z \leq 0.7$, but demonstrate that once the data is binned, the differences between bins can exceed $1\, \sigma$, thus making combining the data questionable. Notably, the overall value masks a descending trend across the bins, the significance of which we find is $2.1 \, \sigma$. Note, our analysis here does not make use of strongly-lensed time delay and is completely independent of H0LiCOW. 

If the H0LiCOW result is substantiated going forward, there are a number of implications. First, this trend cannot be explained by keeping $\Lambda$CDM and adjusting the sound horizon using early Universe physics as in \cite{Poulin:2018cxd, Kreisch:2019yzn, Agrawal:2019lmo}, since this will only raise and lower the trend. Thus, we may be staring at preliminary evidence for a new cosmology at late times. 

Secondly, it is tempting to adopt a different, admittedly speculative, perspective on Hubble tension. Namely, it is conceivable that Hubble tension is still down to two numbers, not the two numbers most consider, namely $H_0 \sim 74$ km s$^{-1}$ Mpc$^{-1}$ (SH0ES) versus $H_0 \sim 67$ km s$^{-1}$ Mpc$^{-1}$ (Planck), but rather the slope and intercept of descending line in $H_0$ with redshift. If true, this provides further hints on the missing piece of the cosmological puzzle, which may be crying out for model building beyond $\Lambda$CDM.

\section{Data}
Let us open with the data. We make use of the following observational results in the redshift range $ z \leq 0.7$: 
\begin{itemize}

\item We employ distances from megamaser hosting galaxies: UGC 3789, NGC 6264, NGC 6323, NGC 5765b, CGCG 074-064 and NGC 4258 in the range $ 0.002 \leq z \leq 0.034$ \cite{Pesce:2020xfe, MaserXI, Reid:2019tiq}.

\item We include cosmic chronometer (CC) data from \cite{CC1, CC2, CC3, CC4, CC5, CC6, CC7}, restricted to the range of interest. 
 
\item Our BAO data comprises isotropic measurements by the 6dF survey ($z = 0.106$) \cite{iso106}, SDSS-MGS survey ($z = 0.15$) \cite{iso115}, as well as the anisotropic measurements by BOSS-DR12 at $z = 0.38, 0.51, 0.61$ \cite{BOSS}. 

\item We incorporate 924 Type Ia SNe from the Pantheon dataset in the range $ 0.01 < z \leq 0.7$ \cite{Scolnic:2017caz}, including both the statistical and systematic uncertainties. 

\end{itemize}

Our overall dataset here is similar to \cite{Dutta:2019pio}, but differs in a number of aspects. Firstly, and most obviously, we have cut the higher redshift data $ z > 0.7$. Secondly, instead of three masers, we now have access to six.  Note, in contrast to \cite{Dutta:2019pio}, following \cite{Pesce:2020xfe} we allow for errors in redshift and corrections for peculiar velocities. We have removed the strong lensing time-delay measurements by H0LiCOW \cite{3lens} to facilitate an independent comparison.  Since we will later bin the data, we have replaced the \textit{compressed} Pantheon data in terms of $H(z)/H_0$ \cite{Riess:2017lxs} (see also \cite{Gomez-Valent:2018hwc}), which was originally considered in \cite{Dutta:2019pio}, by the full dataset \cite{Scolnic:2017caz}.  On the negative side, we have dropped the measurements of $f \sigma_8$, but the omission of this data is not expected to change the conclusions. 

\section{Methodology} 
Any trend in $H_0$ with redshift is model-dependent. Here we focus on flat $\Lambda$CDM, which is described by two parameters: the Hubble constant $H_0$ and matter density $\Omega_m$. We note that CC data is expressed in terms of the Hubble parameter directly, so one can easily fit the model. For megamasers, the relevant distance is the angular diameter distance $D_{A}(z)$, which can be approximated as \cite{Pesce:2020xfe}, 
\be
D_{A} \approx \frac{c z}{H_0 (1+z)} \left(1 - \frac{3 \Omega_m z}{4} + \frac{\Omega_m (9 \Omega_m - 4)z^2}{8} \right). 
\ee
Following \cite{Pesce:2020xfe} we convert between velocities and redshift $v = c z$, allow for an inflated error in the velocities to take into account uncertainties in peculiar velocities $\sigma_{\textrm{pec}} = 250$ km s$^{-1}$ and extremize the following function: 
\bea
\chi^2 = \sum_{i=1}^6 \left[ \frac{(v_i - \hat{v}_i)^2}{\sigma_{v, i}^2 + \sigma_{\textrm{pec}}^2} + \frac{(D({v_i}/{c}) - \hat{D}_i)}{\sigma^2_{D,i}} \right], 
\eea
where the velocities $v_i$ are treated as nuisance parameters and $\hat{v}_i, \hat{D}_i, \sigma_{v, i}, \sigma_{D,i}$ denote the velocities and galaxy distances inferred from modeling maser disks \cite{Pesce:2020xfe}. 

For SNe, as is common practice, we fit the distance modulus 
\be
\mu = m- M = 25 + 5 \log_{10} \left( \frac{D_{L}(z)}{\textrm{Mpc}} \right), 
\ee
where $m$ is the apparent magnitude, $M$ is the absolute magnitude - expected to be $M \approx -19.3$ - that we treat as a fitting parameter, and $D_{L}(z)$ is the luminosity distance, 
\be
D_{L}(z) \equiv c (1+z) \int_{0}^{z} \frac{\dd z'}{H(z')}. 
\ee 

The BAO data involves fitting the following cosmological distances,  
\bea
D_{A}(z) &\equiv& \frac{D_L(z)}{(1+z)^2}, \quad D_{H}(z) \equiv \frac{c}{H(z)}, \nn 
D_{V} (z) &\equiv& [ (1+z) D_{A} (z) ]^{\frac{2}{3}} [ z D_H(z)]^{\frac{1}{3}}. 
\eea
It is important to note that BAO actually measures these quantities divided by  $r_d$ and is only sensitive to the product $r_d H(z)$, which forms the crux of arguments for an early Universe solution to the Hubble tension. In essence, with BAO data fixed, a higher value of $H_0$ requires a lower value of $r_d$ and points to some missing physics before recombination that would reduce this length scale. We refer the reader to  \cite{Evslin:2017qdn} for further discussion on this degeneracy. 

To be fully transparent, it is worth noting that we fit all parameters subject to the following flat priors \footnote{Throughout we employ standard notation for an interval in mathematics.}: 
\be
H_0 \in (0, 100), ~~ \Omega_m \in (0, 1), ~~ r_d \in (0, 200), ~~ M \in (-50, 0). \nonumber
\ee

\begin{figure}[htb]
\centering
\includegraphics[width=90mm]{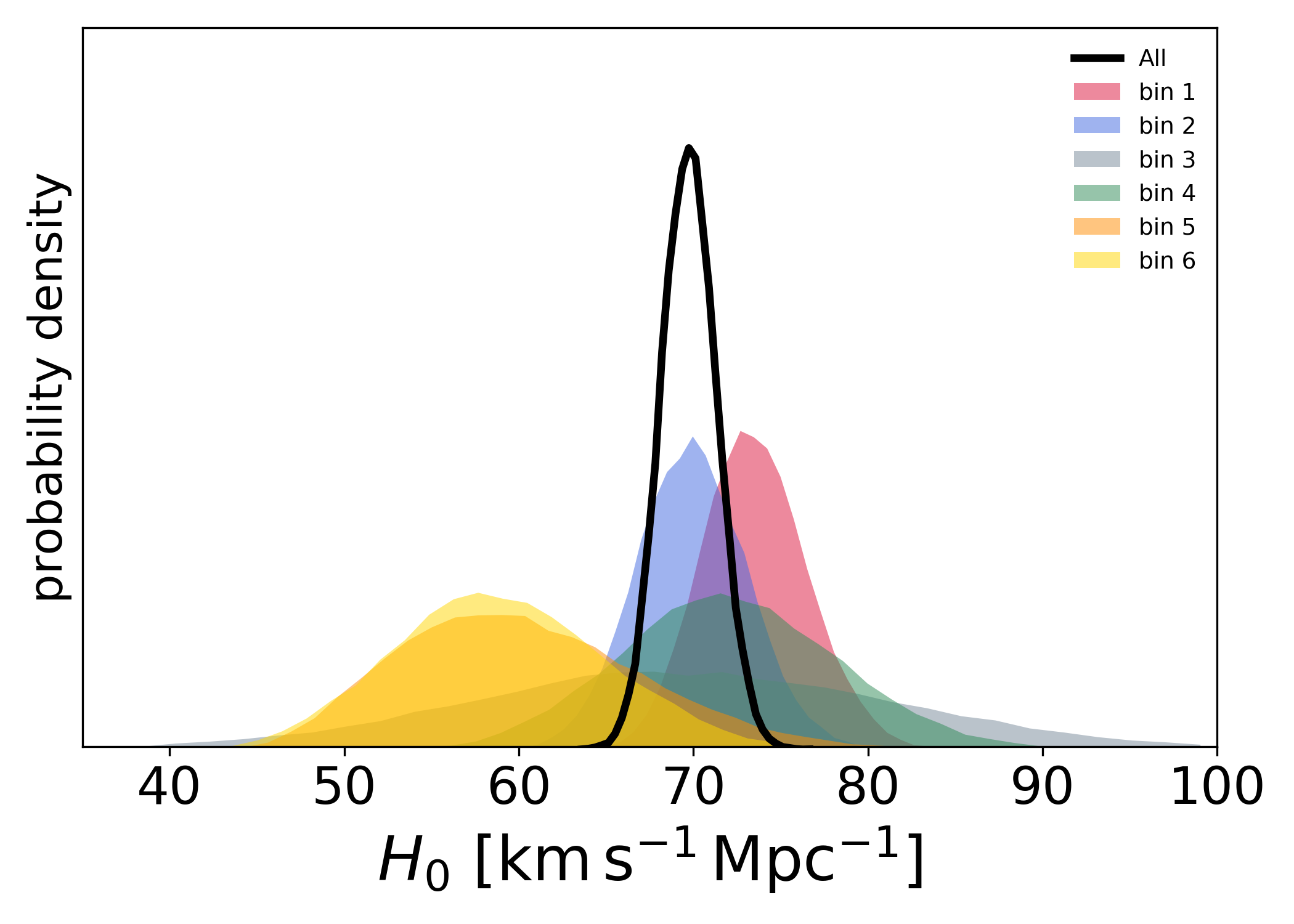}
\caption{Overlapping probability distribution functions for each bin plotted alongside the overall constraint.}
\end{figure}

\section{Binning} 
Overall the dataset is of mixed quality and becomes sparse at higher redshift. Here we employ a non-uniform binning strategy that is designed to achieve a number of results. Neglecting the CC data, which is not so constraining and sparse relative to Type Ia SNe, {but is important in lifting a degeneracy between BAO and SNe alone}, we construct the bins so that the weighted average redshifts of masers, SNe and BAO coincide.  At the same time, we attempt to ensure that the data in a bin is sufficiently constraining and importantly that no data is omitted below $z \leq 0.7$.  
\begin{table}[htb]
\centering
\begin{TAB}(r, 0.3cm, 0.3cm)[5pt]{|c|c|}{|c|cccc|}
Bin & Data \\
1 & Masers, SNe \\
2 & iso BAO, SNe, CC \\
3 & SNe, CC \\
4-6 & aniso BAO, SNe, CC \\ 
\end{TAB} 
\caption{Summary of the data in each bin.}\label{table1}
\end{table}

More concretely, observe that we can define a weighted average redshift of the cosmological probes in a given bin: 
\be
\bar{z}_i = \frac{\sum_k^{N_i} z_k (\sigma_k)^{-2}}{\sum_{k}^{N_i} (\sigma_k)^{-2} }, 
\ee
where $\sigma_k$ denotes the error in the observable at redshift $z_k$. Our strategy is simply to construct bins so that $\bar{z}_i$ for a given data type in that bin coincide, thus allowing us to assign a definite redshift to each bin. 

To this end, we can start from $z = 0.7$ and work backwards in redshift. The upper cut-off is a nominal value, but cannot be much greater than this value as otherwise the BAO data at $z=0.51$ and $z=0.61$ gets binned together. This strategy quickly leads to three bins: 
\bea
\textrm{bin 4:} \quad \bar{z}_4 &=& 0.38 \in (0.321, 0.47],  \nn
\textrm{bin 5:} \quad \bar{z}_5 &=& 0.51 \in (0.47, 0.557],  \nn
\textrm{bin 6:} \quad \bar{z}_6 &=& 0.61 \in (0.557, 0.7],  
\eea
where we have denoted the weighted average value in each bin. By construction the redshifts coincide with BAO. 

To get the first, second and third bin, we identify the weighted average for the masers using $\sigma_k^2 = \sigma^2_{v,k} + \sigma^2_{\textrm{pec}} + \sigma^2_{D, k}$ \cite{Pesce:2020xfe}, which includes an inflated error due to peculiar velocities $\sigma_{\textrm{pec}} = 250$ km s$^{-1}$. This contribution is important as it brings the weighted average redshift into a range where SNe data exists. We are motivated to put the isotropic BAO data at $z = 0.106$ and $z=0.15$ in the same bin to improve the constraining power, reduce the overall number of bins and ensure that the two data points mirror anisotropic BAO data, which at each redshift is also two data points. The remaining SNe we allocate to the final redshift range.
Following the outlined procedure, the remaining bins are 
\bea
\textrm{bin 1:} \quad \bar{z}_{1} &=& 0.021 \in (0, 0.029],  \nn
\textrm{bin 2:} \quad \bar{z}_{2} &=& 0.122 \in (0.029, 0.21], \nn
\textrm{bin 3:} \quad \bar{z}_{3} &=& 0.261 \in (0.21, 0.321]. 
\eea

While this leads to non-uniform bins, and a bin where SNe and CC appear alone, we emphasise again that this way we can confidently assign a definite redshift to each bin. The binning is summarised in Table \ref{table1}.

\section{Results} 
Having discussed the preliminaries, we come to the results. First and foremost, employing the Python package \textit{emcee} \cite{ForemanMackey:2012ig}, we identify the best-fit for the four parameters for the entire dataset through Markov Chain Monte Carlo (MCMC) \footnote{In this analysis, as in \cite{Pesce:2020xfe}, the galaxy velocities are treated as nuisance parameters. That being said, it is heartening to see that their best-fit values from MCMC, namely $3451 \pm 193, 10170 \pm 245, 7796 \pm 248, 8395 \pm 219, 7022 \pm 229, 530 \pm 14$ km s$^{-1}$ agree well with the corresponding values from maser disk modeling $3319 \pm 0.8, 10192.6 \pm 0.8, 7801.5 \pm 1.5, 8525.7 \pm 0.7, 7172.2 \pm 1.9, 679.3 \pm 0.4$ km s$^{-1}$ \cite{Pesce:2020xfe}, once an uncertainty of $250$ km s$^{-1}$ to treat peculiar velocities is taken into account. Overall, this is an indication that our MCMC code is performing as expected.}. The outcome is illustrated in Table \ref{tableII}, confirming that an intermediate value of $H_0 \sim 70$ km s$^{-1}$ Mpc$^{-1}$ is preferred \cite{Dutta:2019pio}. Moreover, $\Omega_m$ and $r_d$ agree with Planck values, $\Omega_m = 0.315 \pm 0.007, r_d = 147.09 \pm 0.26$ Mpc. 

\begin{table}[htb]
\centering
\begin{TAB}(r, 0.5cm, 0.5cm)[5pt]{cccc}{c|c}
$H_0$ [$\frac{\textrm{km}}{\textrm{s Mpc}}$] & $\Omega_m$ & $r_d$ [Mpc] & $M$ \\
$69.74^{+1.60}_{-1.56}$ & $0.30^{+0.02}_{-0.02}$ & $144.83^{+3.44}_{-3.34}$ & $-19.36^{+0.05}_{-0.05}$ \\
\end{TAB} 
\caption{Best-fit values of the maser+CC+SNe+BAO dataset over the redshift range $ z \leq 0.7$.}\label{tableII}
\end{table}

Following similar analysis, but tailoring the MCMC to the data in the bin, we identify the best-fit values for the parameters in each bin as shown in Table \ref{tableIII} {and illustrated in Figure 1}. As is evident, there is a trend whereby $H_0$ decreases with redshift. This is primarily down to the higher $H_0$ value coming from the masers in the first bin, but the anisotropic BAO data is also playing a role in driving $H_0$ lower. In the process, the best-fit values for $r_d$ and $M$ start to drift outside of $1 \, \sigma$ of the canonical values. Note, we have imposed no assumptions and it is simply data that is guiding us in this direction. 

\begin{figure}[htb]
\centering
\includegraphics[width=90mm]{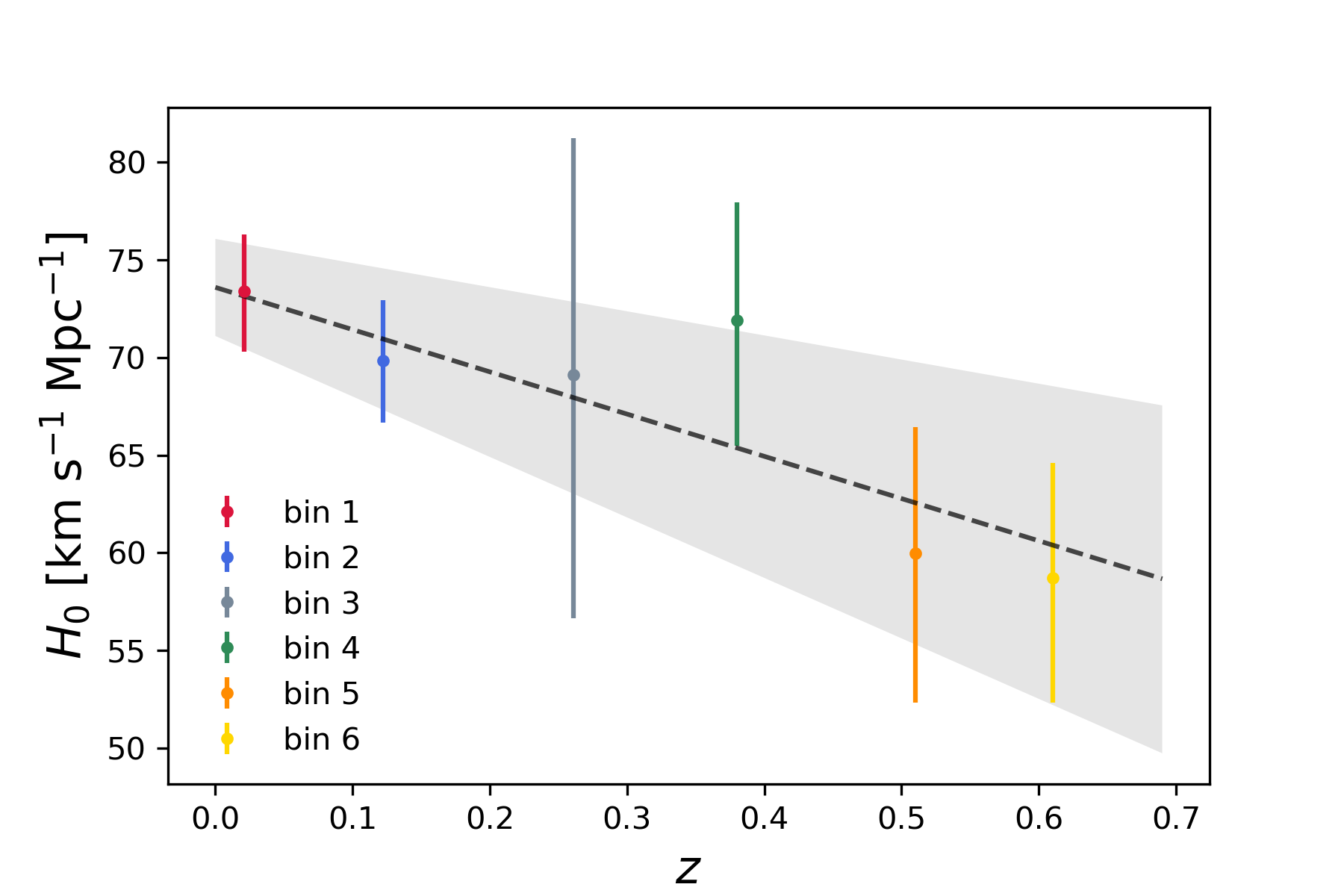}
\caption{The best-fit line against the binned data. The line is $2.1 \, \sigma$ removed from the flat line null hypothesis.}\label{Fig2}
\end{figure}

To get a handle on the significance, we follow earlier H0LiCOW analysis \cite{Wong:2019kwg} to establish a null hypothesis. This is done by shifting the probability distribution function (pdf) for each bin to the central value favoured by the complete dataset $H_0 = 69.74$, before drawing a set of six mock $H_0$ values using the respective pdfs. Once this is done, we perform a weighted fit to identify a line and repeat $10^5$ times. Doing so, one will find a normal distribution peaked on a zero slope from where one can infer confidence intervals. We fit the same linear regression through the data with the original \textit{binned} $H_0$ values and find that the slope of the data falls $2.1 \, \sigma$ away from the slope of the null hypothesis.  Concretely, we find the best-fit line has slope $m = -21.7 \pm 9.4 $ with intercept $H_0 = 73.6 \pm 2.5$, where the intercept is curiously close to H0LiCOW's $H_0$ determination \cite{Wong:2019kwg}. We illustrate it against the binned data in Figure \ref{Fig2}.

\begin{table}[htb]
\centering
\begin{TAB}(r, 0.3cm, 0.3cm)[5pt]{ccccc}{c|cccccc}
$\bar{z}$ & $H_0$ [$\frac{\textrm{km}}{\textrm{s Mpc}}$]  & $\Omega_m$ & $r_d$ [Mpc] & $M$ \\
$0.021$ & $73.41^{+3.10}_{-2.88}$ & $0.51^{+0.33}_{-0.34}$ & $-$ & $-19.26^{+0.09}_{-0.09}$ \\
$0.122$ & $69.85^{+3.17}_{-3.10}$ & $0.26^{+0.10}_{-0.09}$ & $143.08^{+7.14}_{-6.74}$ & $-19.36^{+0.09}_{-0.09}$ \\
$0.261$ & $69.10^{+12.46}_{-12.12}$ & $0.27^{+0.20}_{-0.15}$ & $-$ & $-19.39^{+0.40}_{-0.33}$ \\
$0.38$ & $71.90^{+6.42}_{-6.03}$ & $0.22^{+0.11}_{-0.09}$ & $143.94^{+9.94}_{-8.91}$ & $-19.33^{+0.15}_{-0.15}$ \\
$0.51$ & $59.98^{+7.64}_{-6.45}$ & $0.37^{+0.12}_{-0.10}$ & $164.05^{+17.66}_{-15.92}$ & $-19.65^{+0.23}_{-0.23}$ \\
$0.61$ & $58.72^{+6.40}_{-5.87}$ & $0.44^{+0.12}_{-0.10}$ & $161.04^{+13.31}_{-11.55}$ & $-19.59^{+0.18}_{-0.17}$ \\
\end{TAB} 
\caption{Neglecting the velocity nuisance parameters in the first bin, we record the best-fit values for the Hubble constant, matter density, the sound horizon radius and absolute magnitude of Type Ia SNe in the remaining bins.}\label{tableIII}
\end{table}

\section{Discussion}
H0LiCOW have reported a descending trend of measured $H_0$ with lens redshift, which is not the result of any obvious systematic. The deviation from a horizontal line is currently $1.7 \, \sigma$. In this letter using a combined dataset of masers, SNe, BAO and CC, which overall favour a central value for $H_0$, we have provided independent evidence for such a trend in a similar redshift range with statistical significance $2.1 \, \sigma$. While our slope is consistent with H0LiCOW, there is a difference in the intercept. That being said, it should be borne in mind that the underlying data is different in nature.  

On the robustness of the result, it is worth noting that the correlation is driven by higher values of $r_d$ relative to Planck ($r_d \approx 147$ Mpc) due to CC data in both bins 5 and 6. In other words, removing one of the bins will not make a difference, but as is clear from Figure \ref{Fig2} omitting both bin 5 and bin 6 will remove the descending trend.} Alternatively, one can start eliminating datasets. Removing SNe from the analysis does not change the result, while removing BAO leaves our analysis resting on CC data, which only inflates the error bars so that a horizontal line can be fitted. We cannot remove CC as it is instrumental in breaking the degeneracies from both BAO and SNe. It is worth noting that bins 4, 5 and 6 employ the same combination of data, as is clear from Table \ref{table1}. So, it may be puzzling that the $H_0$ values differ considerably in Figure \ref{Fig2}. However, here the number of CC data points respectively in bins 5 and 6 are two, which can be contrasted with seven CC data points in bin 4. Thus, the quality of the CC data in the latter bins is allowing $r_d$ greater freedom. This explains the difference.

Once again we reiterate that we have not assumed a prior on $r_d$ and have let the data do the talking. This is essentially to ensure our analysis only depends on late Universe physics. Nonetheless, we have checked that if one adopts a Planck prior on $r_d$, the significance of the line decreases to $1.4 \, \sigma$ and further decreasing $r_d$ to the values favoured by the EDE proposal, the binned values of $H_0$ are fully consistent with a horizontal line. Note, there is a tautological quality to the latter. Since BAO strongly constrains fitting, adopting a prior on $r_d$ is tantamount to fixing $H_0$ from the outset, which is precisely what we wanted to avoid.

Finally, one may be concerned that $H_0$ in the last two bins, namely $H_0 \sim 60$ km s$^{-1}$ Mpc$^{-1}$, is too low compared to Planck. Clearly, since H0LiCOW reports higher than expected values, i. e. $H_0 \sim 80$ km s$^{-1}$ Mpc$^{-1}$, at lower lens redshift, as observed, there is a discernible difference in the intercept. However, just as we have high and low values in bins, but the overall best-fit  is a central value (Table II), it is conceivable that the Planck result for flat $\Lambda$CDM is an ``averaged" value, which is essentially a \emph{coarse-grained} value for $H_0$.

In the big picture, provided the descending trend reported by H0LiCOW can be substantiated in future, it will call into question early Universe solutions to the Hubble tension. Put simply, our analysis here is exploiting CC data, which in contrast to BAO is model independent, to fix $r_d$: we have not imposed a prior. Moreover, this is also true in the H0LiCOW analysis, where $r_d$ does not even appear. Therefore, the descending feature in $H_0$ that we and/or H0LiCOW have found is difficult to explain by fiddling with the length scale $r_d$ while keeping $\Lambda$CDM intact, since this will simply raise and lower the trend, but will not remove it. Of course, such a feature requires explanation and we will return to this in future work.

To whet the appetite, let us present a simple perturbative argument that serves to highlight some issues. The idea is to take the flat $\Lambda$CDM cosmology but with the replacement $H_0 \rightarrow H_0 + m z$. Next, let us equate the corresponding Hubble parameter with one from a new cosmology where the equation of state for dark energy $w_0$ and dark matter $w_c$ are allowed to vary from their $\Lambda$CDM values $w_0 = -1, w_c = 0$ (we assume that $w_c > 0$). Taylor expanding around $z=0$ and making comparison to linear order in redshift tells us that $w_c$ cannot explain a negative slope as it has the wrong sign. Moreover, for canonical values of matter density we find that a descending feature may be explained by $w_0 \approx -1.3$, which is consistent with the naive value of $w_0$ required to resolve the Hubble tension fully. Note to linear order in $z$, the $w_a$ term in the CPL parametrisation \cite{Chevallier:2000qy, Linder:2002et} does not appear, so it cannot change the result even if it was added. However, $w_0 \approx -1.3$ is disfavoured by existing results \cite{Aghanim:2018eyx,Scolnic:2017caz}, so explaining this feature, if real, requires a little more imagination, e.g. see \cite{Liao:2020zko}.

\section*{Acknowledgements}
We thank Lucas Macri, Vivian Poulin, Adam Riess, Dan Scolnic, Sunny Vagnozzi and Kenneth Wong for correspondence. MMShJ would like to thank Nima Khosravi for an earlier discussion on a similar idea. This work was supported in part by the Korea Ministry of Science, ICT \& Future Planning, Gyeongsangbuk-do and Pohang City. MMShJ would like to thank the hospitality of ICTP EAIFR where this research carried out and acknowledges the support by 
INSF grant No 950124 and Saramadan grant No. ISEF/M/98204.
Ruchika acknowledge the funding from CSIR, Govt. of
India under Senior Research Fellowship. AAS and CK acknowledge APCTP, Pohang, Korea for supporting their visits during the workshop ``APCTP lecture series on (evidence for) physics beyond Lambda-CDM" where this work was initiated.

\end{document}